\documentclass[twocolumn,nofootinbib,aps,preprintnumbers,amsmath,amssymb]{revtex4-1}
\usepackage{amsmath}
\usepackage{graphicx}
\usepackage{dcolumn}
\usepackage{bm}
\usepackage{epsfig,color,xspace,bbold}
\usepackage{setspace}
\usepackage{url}
\usepackage[latin9]{inputenc}                                                             
\usepackage{graphicx}                                                                     
\usepackage{amsmath}                                                                      
\usepackage{amssymb}                                                                      
\usepackage{rotating}

\usepackage[version=3]{mhchem} 

\begin{document}
\title{Big Data meets Quantum Chemistry Approximations:                                   
The $\Delta$-Machine Learning Approach}   

\author{Raghunathan Ramakrishnan}
\affiliation{Institute of Physical Chemistry and National Center for Computational Design and Discovery of Novel Materials, 
Department of Chemistry,               
University of Basel, Klingelbergstrasse 80, CH-4056 Basel,                                
Switzerland}

\author{Pavlo O. Dral}
\affiliation{Computer-Chemie-Centrum and Interdisciplinary Center for Molecular 
Materials, Department Chemie und Pharmazie, Friedrich-Alexander-Universit\"at 
Erlangen-N\"urnberg, N\"agelsbachstrasse 25, 91052 Erlangen, Germany}
\affiliation{Max-Planck-Institut f\"ur Kohlenforschung, Kaiser-Wilhelm-Platz 1, 45470 M\"ulheim an der Ruhr, Germany}

\author{Matthias Rupp}
\affiliation{Institute of Physical Chemistry and National Center for Computational Design and Discovery of Novel Materials, 
Department of Chemistry,               
University of Basel, Klingelbergstrasse 80, CH-4056 Basel,                                
Switzerland}

\author{O. Anatole von Lilienfeld}
\affiliation{Institute of Physical Chemistry and National Center for Computational Design and Discovery of Novel Materials, 
Department of Chemistry,               
University of Basel, Klingelbergstrasse 80, CH-4056 Basel,                                
Switzerland}
\affiliation{Argonne Leadership Computing Facility, 
Argonne National Laboratory, 9700 S. Cass Avenue, Lemont, IL 60439, USA}
\email{anatole.vonlilienfeld@unibas.ch}




\begin{abstract}
Chemically accurate and comprehensive studies of the virtual space of all                 
possible molecules are severely limited by the computational cost of quantum chemistry.   
We introduce a composite strategy that adds machine learning                              
corrections to computationally inexpensive approximate legacy quantum methods.            
After training, highly accurate predictions of enthalpies,                                
free energies, entropies, and electron correlation energies are possible,
for significantly larger molecular sets than used for training.                           
For thermochemical properties of up to 16k constitutional isomers of C$_7$H$_{10}$O$_2$
we present numerical evidence that chemical accuracy can be reached.
We also predict electron correlation energy in post Hartree-Fock methods, at
the computational cost of Hartree-Fock, and we 
establish a qualitative relationship between molecular entropy and electron correlation.            
The transferability of our approach is demonstrated, using semi-empirical quantum chemistry
and machine learning models trained on 1 and 10\% of 134k organic molecules,
to reproduce enthalpies of all remaining molecules at density functional theory level of accuracy.
\end{abstract}
\maketitle

\section{Introduction} 
Designing new molecular materials is one of the key challenges in chemistry, 
and a major obstacle in solving many of the pressing issues that today's society faces, 
such as clean and cheap water, advanced energy materials, or novel drugs to fight antibiotic resistant bacteria.
Unfortunately, the number of potentially interesting small molecules is too large for exhaustive 
screening~\cite{ChemicalSpace,BeratanUnchartedCCS2013,anatole-ijqc2013},
even when relying on automated synthesis and combinatorial high-throughput 
``click-chemistry''~\cite{CombinatorialThinking1997,ClickChemistrySharpless2001}.
Virtual screening strategies, made feasible by ever increasing compute power, 
advanced atomistic simulation software, 
and quantitative structure-property relationships have already helped in the discovery 
of new materials, and provided crucial guidance 
for subsequent experimental characterization and 
fabrication~\cite{ReviewCatalystNorskov,JorgensenScience2004,DrugDesignEugeneShakhnovich2010,MGI2011,MGP2013,CederPRB2013}.
To achieve the overall goal of {\em de novo in silico} molecular and materials 
design~\cite{Beratan1996,ZungerNature1999,anatole-prl2005,RCDYang2006}, however,
substantial progress is still necessary~\cite{SchneiderReview2010},
especially regarding prediction accuracy, computational speed, and transferability
of the employed models.

For quantum chemistry models to attain ``chemical accuracy'' ($\thickapprox$ 1~kcal/mol) in
the prediction of covalent binding is crucial in many scientific domains. 
Examples include the understanding of combustion processes~\cite{CombustionChemistry,CombustionIsomersKohse2012,CombustionAccuracy-Green2013}; 
questions relevant interstellar chemistry~\cite{Duleyinterstellarchem};
and prediction of reaction rates essential for catalysis. The latter
depend exponentially on energy differences,
implying that small errors on the order of $k_B T$ propagate dramatically.
More generally, reaching chemical accuracy can be crucial for
the detection of new structure property relationships, trends or patterns in  
Big Data, the design of new molecular materials with sensitive property requirements,
or the energetics of competing reactants and products determining mechanisms and reaction rates. 
Control over the accuracy of important thermochemical properties of molecules can 
be achieved through application of well-established hierarchies in quantum chemistry~\cite{friesner2005ab}.
Calibrated composite methods such as John Pople's Gaussian model chemistry
exploit the inherent transferability of corrections to electronic correlation, 
the Born-Oppenheimer approximation, or basis-set deficiencies~\cite{G1,G2}.
This has enabled chemists to routinely achieve chemical accuracy for 
{\em any} non-exotic and medium-sized organic molecule at substantial yet manageable computational costs~\cite{G4,G4MP2}.

Unfortunately, such calculations are too demanding for the routine investigation of 
larger subsets of chemical space.
Note, however, that the computationally most demanding 
task in a quantum chemistry calculation corresponds to an energy contribution 
that constitutes only a minor fraction of the total energy, while most of the relevant physics 
can already be accounted for through computationally very efficient approximate legacy quantum chemistry, 
such as the semi-empirical theory PM7, Hartree-Fock (HF), or even density functional theory (DFT).
For the water molecule H$_2$O, for example, HF approximates
the experimental ionization potential by more than 90\%~\cite{SzaboOstlund}.
Calculating the remaining $\Delta$ with chemical accuracy using correlated electronic 
structure methods 
requires a disproportionate amount of computational effort due to unfavorable 
pre-factors and scaling with number of electrons. 
In this study, we introduce an alternative {\em Ansatz} to model the expensive $\Delta$ 
using a statistical model trained on reference data requiring only a fraction of the computational cost. 
The observed speed-up, up to several orders of magnitude,
is due to the computational efficiency of machine learning (ML) models. 
We have validated this idea for several molecular properties, 
combining quantum chemistry results at several levels of theory 
with $\Delta$-ML models trained over comprehensive molecular data sets
drawn from 134 kilo organic molecules published in Ref.~\onlinecite{DATAPAPER}.
While the basic idea to augment approximate models with ML is not new~\cite{hwwc2003,basissetBalabin,manybodyCsanyi},
we present a generalized $\Delta$-ML-model that achieves
unprecedented chemical accuracy and transferability.

We present numerical evidence for predicted atomization enthalpies, 
free energies, and electron correlation in many thousands of organic molecules
(reaching molecular weights of up to 150 Dalton) 
with an accuracy of $\thickapprox$1~kcal/mol 
at the computational cost of DFT or PM7.
We validate the $\Delta$-ML model for entirely new subsets of chemical space,
up to two orders of magnitude larger than the set used for training. 
Using $\Delta$-ML-based screening, we  find that within the
constitutional isomers of C$_7$H$_{10}$O$_2$,
molecular entropy and electron correlation energy of atomization 
are not entirely independent from each other. 
This suggests not only significant coupling between electronic and vibrational eigenstates 
but also the existence of Pareto fronts that can impose
severe limitations with respect to simultaneous property optimization.
Finally, we establish transferability by accurately
predicting properties for a much larger molecular dataset comprising of 134k molecules.


\section{The $\Delta$-ML approach}
\begin{figure}[hpt]
\centering                                                  
\includegraphics[width=8.8cm, angle=0.0, scale=1]{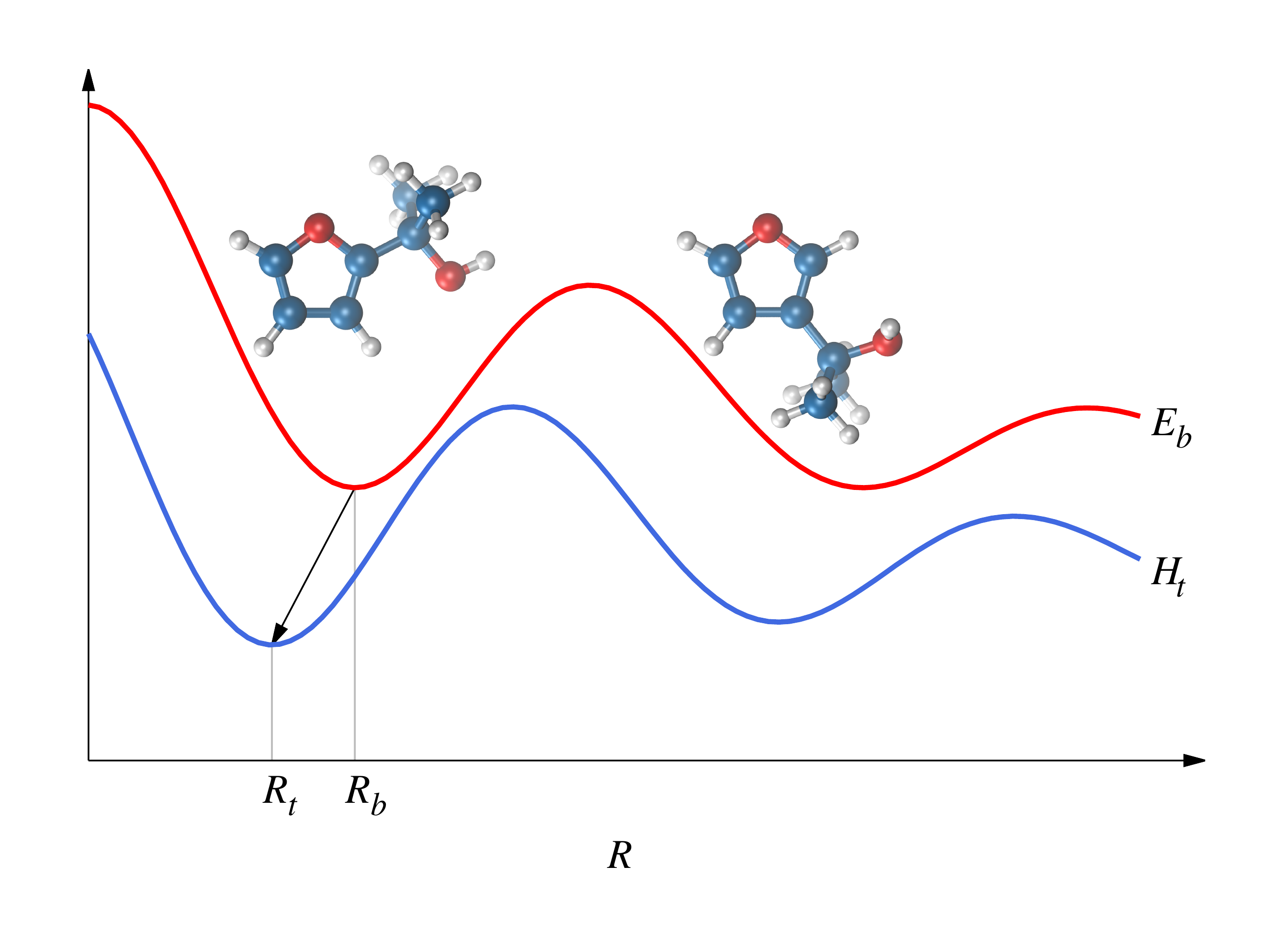}
\caption{
Two hypothetical property profiles 
connecting two constitutional isomers of C$_7$H$_{10}$O$_2$.
The $\Delta$-model, Eq.~(\ref{eq:Model1}), estimates the difference between baseline  and 
targetline properties (arrow) which differ in level of theory ($b \rightarrow t$), 
geometry ($R_b \rightarrow R_t$), and property ($E_b \rightarrow H_t$).} 
\label{fig:delta} 
\end{figure}

The $\Delta^t_b$-model of a molecular property corresponds to 
a baseline ($b$) value plus a correction, towards a targetline ($t$) value, modeled statistically.
More specifically, given a property $P'_b$, such as the energy $E_b$, 
for the relaxed geometry $R_b$ of a new query molecule, calculated using 
an approximate baseline level of theory, 
another related property $P_t$, such as the enthalpy $H_t$, 
corresponding to a more accurate and more demanding target level of theory can be estimated as
\begin{equation}
P_t (R_t) \; \approx \; \Delta^t_b(R_b) \; = \; P'_b(R_b) + \sum_{i=1}^N \alpha_i k(R_b,R_i).
\label{eq:Model1}
\end{equation}
The sum represents an ML-model, here a linear combination of Slater-type basis functions,
$k(R_b,R_i) = e^{-|R_i-R_b|/\sigma}$, centered on $N$ training molecules,
and with global hyperparameter $\sigma$---the kernel's width.
The regression coefficients $\{\alpha_i\}$ are obtained through kernel ridge regression, a regularized nonlinear 
 regression model~\cite{htf2009} that limits the norm of regression coefficients,
thereby reducing overfitting and improving the transferability of the model to new molecules.
$|R_i-R_b|$ is a quantitative measure of similarity between query molecule and training molecule $i$,
using the Manhattan-norm ($L_1$) between sorted Coulomb matrix representations~\cite{RuppPRL2012,AssessmentMLJCTC2013}.
The latter uniquely encodes (except among enantiomers) 
the external potential of any
given molecule in a way that is invariant with respect to molecular
translation, rotation, or atom-indexing.
As such $P_t$ of a new molecule, consistent with its minimum geometry~$R_t$ at the target level of theory, 
is estimated using exclusively $R_b$ and $P'_b$ as input.
Thus, the $\Delta$-model accounts for differences in 
(i) definition of property observable, e.g. energy $\rightarrow$ enthalpy, 
(ii) level of theory, e.g. PM7 $\rightarrow$ G4MP2, and 
(iii) changes in geometry (illustrated in \ref{fig:delta}).

As a first test of our {\em Ansatz}, we have trained $\Delta_b^t$ models for HOMO and LUMO
eigenvalues calculated at various levels 
of theory~\cite{Montavon2013} for the smallest 7k organic molecules in the GDB-dataset introduced
by Reymond and coworkers~\cite{GDB17}.
After training on calculated data for 1k molecules, the resulting "1k $\Delta$-model" reduces the mean absolute error (MAE) 
in the prediction of GW HOMO eigenvalues for the remaining 6k molecules from 0.78 to 0.23 eV
for the semi-empirical ZINDO baseline method.
Interestingly, while the less empirical DFT hybrid (PBE0) baseline method has an MAE of more than 2 eV, 
this reduces to less than 0.1 eV when combined with the 1k $\Delta$-model. 
Correspondingly, MAEs for predicting GW LUMO eigenvalues reduce from 0.91 to 0.16 eV and from 1.3 to 0.13 eV 
for $\Delta_{\rm ZINDO}^{\rm GW}$ and $\Delta_{\rm PBE0}^{\rm GW}$, respectively.
This observation suggests that more 
sophisticated baseline models, albeit occasionally leading to more substantial
errors than simpler models, overall are smoother in chemical space, 
and therefore easier to learn.

\begin{figure}[hpt]
\centering                                                  
\includegraphics[width=8.8cm, angle=0.0, scale=1]{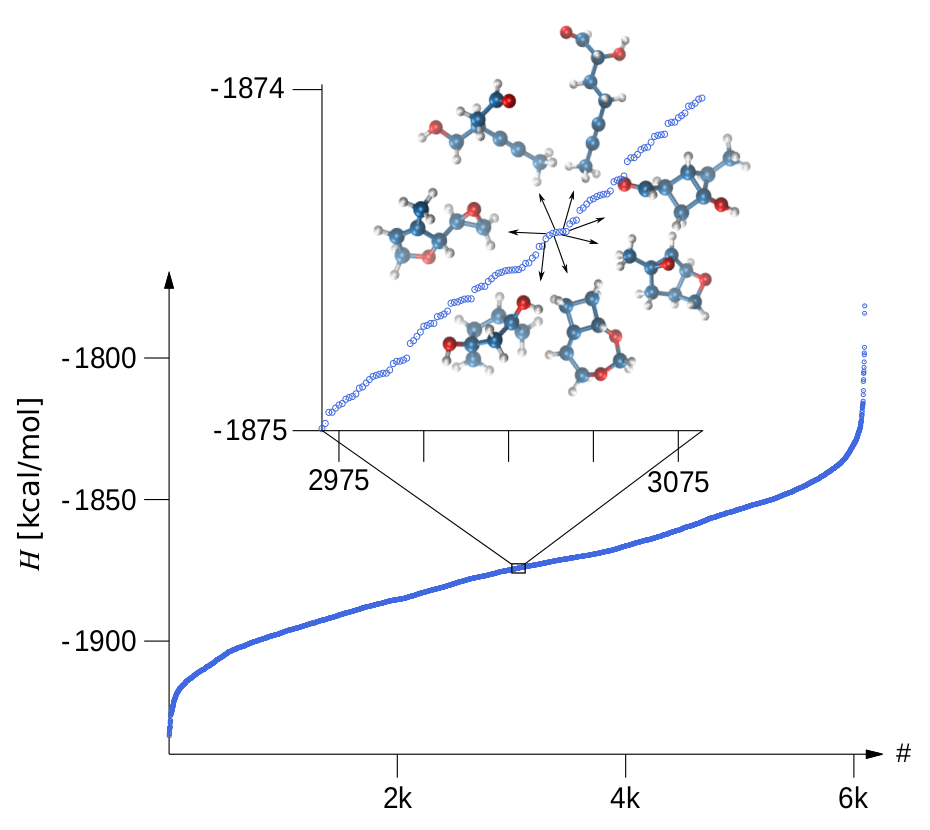}
\caption{
Illustration of chemical diversity and data density 
of up to $\sim$100 molecules per kcal/mol of atomization enthalpy $H$ 
(G4MP2 level of theory~\cite{G4,G4MP2}),  shown in ascending order for all 6k constitutional isomers 
of C$_7$H$_{10}$O$_2$ in the GDB-17 data-set~\cite{GDB17,DATAPAPER}.
Seven near degenerate (within $\approx$ 0.01 kcal/mol) molecules in the inset exemplify the chemical 
diversity.
}                                                                                         
\label{fig:density}                                                                          
\end{figure} 

%
%

\section{Results and Discussion}
\subsection{Chemically accurate prediction of covalent bonding}
To demonstrate that the $\Delta$-ML model can reach chemical accuracy, 
we have investigated the covalent binding energies of the 6k constitutional isomers of
C$_7$H$_{10}$O$_2$ in the GDB~\cite{GDB17}.
Note that any other molecular set could have been used just as well.
We relied on previously calculated highly accurate target level atomization energies ($E_t$ in \ref{fig:delta}) 
at the G4MP2 level \cite{G4MP2} for these isomers~\cite{DATAPAPER}. 
G4MP2 is widely considered to be on par with experimental uncertainties~\cite{DFTreview-CohenYang2012cr}.
While structurally highly diverse, this data-set exhibits many near-degeneracies 
with high energy densities of up to $\sim$100 molecules per kcal/mol in atomization enthalpy $H$
( inset of \ref{fig:density}).

\begin{figure}[hpt]
\centering                                                  
\includegraphics[width=8.8cm, angle=0.0, scale=1]{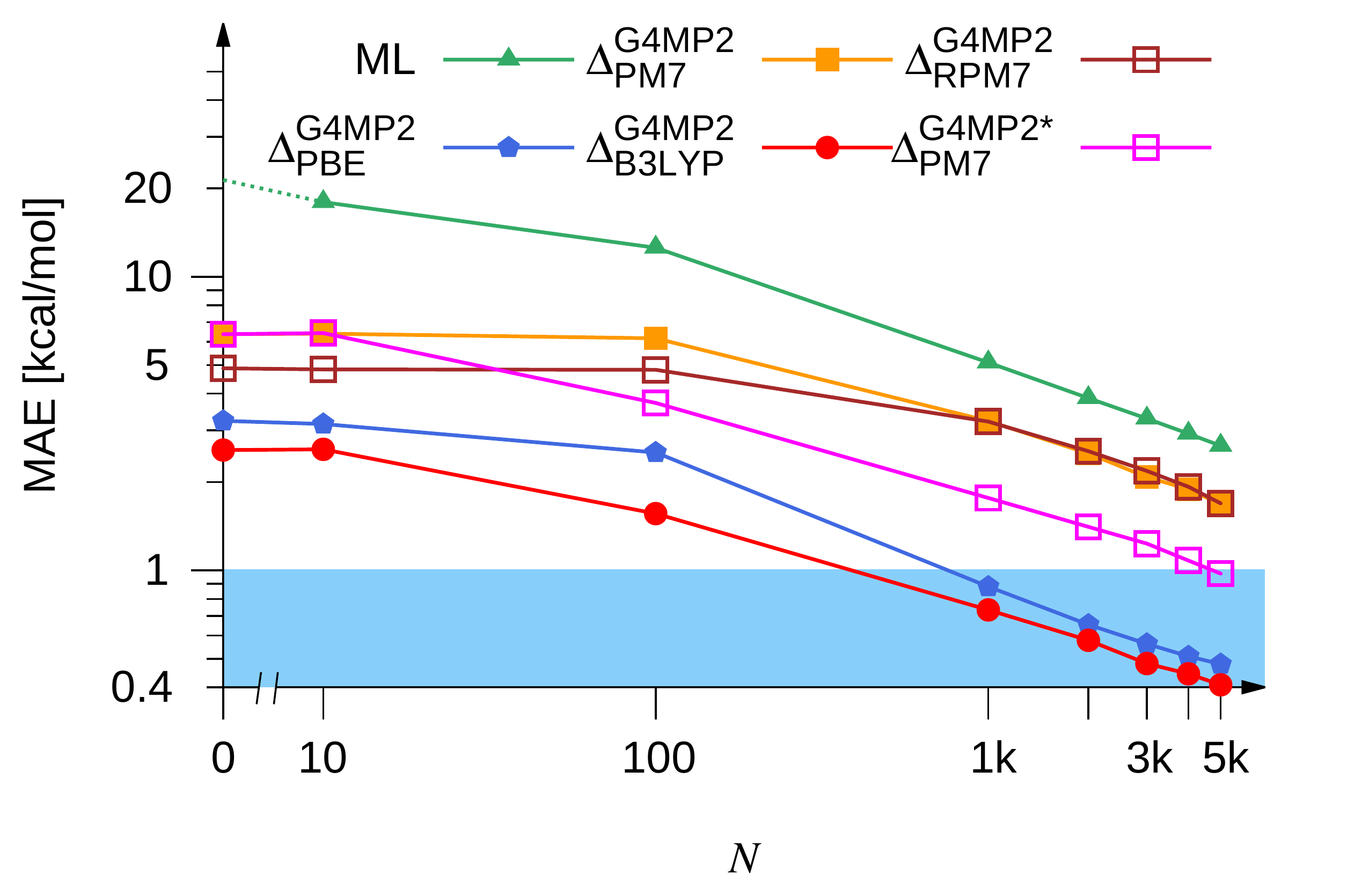}
\caption{                                                                                 
Mean absolute error (MAE) [kcal/mol] of ML predicted atomization energies compared to G4MP2 reference values
as a function of training set size $N$ for out-of-sample predictions.
The lines correspond to various baselines in the $\Delta^{\rm G4MP2}_b$-model (Eq.~[\ref{eq:Model1}]). 
The MAE at $N$ = 0 represents the baseline's error.
For comparison, the baseline-free ML model is shown as well (green); 
its $N$=0 value is the standard deviation in the G4MP2 atomization energies of the data set.
The "chemical accuracy" target  of 1 kcal/mol is highlighted in blue.
$\Delta^{\rm G4MP2}_{\rm RPM7}$ (brown) and $\Delta^{\rm G4MP2^*}_{\rm PM7}$ (pink) are
variants of $\Delta^{\rm G4MP2}_{\rm PM7}$  (yellow) using reparameterized
PM7 as baseline or an alternative molecular representation, respectively. 
}                                                                                         
\label{fig:curve}                                                                          
\end{figure}

The  $\Delta$-ML model's  
systematic improvement of accuracy with increasing training set size 
is shown as a log-log-plot in \ref{fig:curve} for the potential energy of atomization. 
Starting at different offsets, corresponding to the respective error of the pure baseline methods, 
the MAE, measured out-of-sample on the remaining molecules in the 6k set,
rapidly decreases.
Note the constant decay rate for training set sizes larger than 1000 for 
all $\Delta_b^{\rm G4MP2}$-ML model errors. 
This suggests that for all models the error could be lowered even further 
if more training data were used.

While the error of the GGA DFT baseline model ($\Delta_{\rm PBE}^{\rm G4MP2}$) 
in \ref{fig:curve}
starts off slightly higher than the more accurate hybrid DFT analogue ($\Delta_{\rm B3LYP}^{\rm G4MP2}$),
both rapidly converge to chemical accuracy ($<$ 1~kcal/mol) for less than 1k training molecules.
$\Delta_{\rm B3LYP}^{\rm G4MP2}$ reaches $\sim$0.4~kcal/mol for a 5k training set. 
For the substantially faster, yet more approximate PM7 baseline model 
($\Delta_{\rm PM7}^{\rm G4MP2}$) an accuracy similar to pure hybrid DFT (B3LYP)
can be achieved with 2k training molecules, and for 5k training molecules 
the error has been quenched to less than 2~kcal/mol.
All $\Delta$-ML models outperform an  ML model
trained on the absolute value of $E_t$ directly~\cite{RuppPRL2012,AssessmentMLJCTC2013}, without a baseline.

We also considered the  effect of reparameterizing the baseline method
on the training set before applying $\Delta$-ML.
The arguably simplest model, Benson's thermochemical bond additivity model
(bound counting), has a prediction error of MAE$\thickapprox100$ kcal/mol for the 6k isomers. 
After reparameterization of all bond energies to fit the training data, its MAE reduces to
$\thickapprox30$ kcal/mol, which is worse than direct ML.
Along the same line, we optimized all semi-empirical parameters in PM7 in order to reproduce 
G4MP2 atomization enthalpies of up to 128 C$_7$H$_{10}$O$_2$ isomers, drawn at random. 
\ref{fig:curve} (brown line) shows that when using this reparameterized 
(RPM7) baseline model in $\Delta_{\rm RPM7}^{\rm G4MP2}$
it does allow for an improved off-set, however the advantage 
 vanishes when increasing training set size beyond 1k (\ref{fig:curve}).  
Using the semi-empirical model OM2~\cite{weber2000orthogonalization}, 
instead of PM7, we found a similarly vanishing effect. 
These findings indicate the severe limitations inherent in fixed functional 
forms of electronic semi-empirical model Hamiltonians used in combination with globally
optimized parameters---no matter the actual combination of parameters. 
Statistically learned $\Delta$-corrections, inferred from large numbers of example molecules, 
however, seem capable of capturing the more delicate energy contributions in the 
G4MP2 energy. 
We have also tested the effect of using an alternative
molecular representation. 
The $\Delta_{\rm PM7}^{\rm G4MP2^*}$ model in \ref{fig:curve} (pink line)
shows the improvement of performance of the $\Delta_{\rm PM7}^{\rm G4MP2}$ model
when replacing the above mentioned Coulomb-matrix representation by the
bag-of-bond descriptor recently introduced by one of us~\cite{BobPaper}.
Encouragingly, also for this descriptor one observes similar decay rates,
and an even better performance than for the Coulomb-matrix based $\Delta$-model,
reaching chemical accuracy  for a training set size of 5k.

\subsection{Chemically accurate thermochemistry}
Prediction accuracy for thermochemical properties,
such as internal energies, enthalpies, free energies and entropies of atomization at
298.15 K,
all trained to reproduce G4MP2 target level of theory
for the same set of 6k constitutional isomers of C$_7$H$_{10}$O$_2$ were investigated. 
$\Delta$-ML models have been trained for three baselines,
$\Delta^{\rm G4MP2}_{\rm PM7}, \Delta^{\rm G4MP2}_{\rm PBE}$, and $\Delta^{\rm G4MP2}_{\rm B3LYP}$, 
on subsets of varying sizes. 
The baseline properties corresponded in this case simply 
to the potential energy of atomization, with the ML model accounting for differences
in level of theory, in geometry, as well as for the respective thermodynamic effects. 
\ref{tab:Thermo} lists  resulting errors and standard deviations  of 
predicted enthalpies of atomization at 298.15 K for various trainingset sizes. As before,
$\Delta$-ML  models 
display rapid error decay with increasing training set size.
Encouragingly the standard deviation also decays rapidly with training set size.
Again, the DFT baseline models yield MAEs of less than 1~kcal/mol already at 1k training set size,
and the error of the 5k-$\Delta_{\rm B3LYP}^{\rm G4MP2}$-ML model remains
below even after addition of the standard deviation. 
The computationally less expensive PM7 baseline model performs slightly worse than the DFT based models.
The 1k-$\Delta_{\rm PM7}^{\rm G4MP2}$-ML model decreases the pure PM7 prediction error and standard
deviation by more than $\sim$50\%, and converges to near chemical accuracy (1.7~kcal/mol) 
for a 5k training set. Computational effort for out-of-sample predictions is dominated by
baseline evaluations.
For internal energies, free energies and entropies of atomization 
we have observed nearly identical convergence and baseline trends. 
All these results indicate that the $\Delta$-ML approach represents an 
inexpensive strategy to accurately estimate not only differences in 
potential energies due to different electronic structure models,
but to also account for thermal contributions to 
thermodynamic state functions {\em without} having to calculate the 
corresponding partition functions. 
Note that the latter can be prohibitively expensive when using more accurate theories.

\begin{table}[hpb]
\caption{                                                                                
Mean absolute errors $\pm$ standard deviations for predicted out-of-sample enthalpies of 
atomization $H$ ($T$=298.15 K) at G4MP2 level of theory using 
the $\Delta_b^{\mathrm{G4MP2}}$-ML model
for increasing training set size $N$ in Eq.~(\ref{eq:Model1}).
All values in~kcal/mol.
Training and test set sizes always add up to 6095 constitutional isomers of C$_7$H$_{10}$O$_2$.
}
\begin{tabular} {@{\vrule height 10.5pt depth4pt  width0pt}l|r@{$\pm$}l|r@{$\pm$}l|r@{$\pm$}l}
$N$ & \multicolumn{2}{c}{$\triangle_{\mathrm{PM7}}^{\mathrm{G4MP2}}$} & \multicolumn{2}{c}{$\triangle_{\mathrm{PBE}}^{\mathrm{G4MP2}}$} & \multicolumn{2}{c}{$\triangle_{\mathrm{B3LYP}}^{\mathrm{G4MP2}}$} \\\hline 
0  & 6.4&8.6 & 3.0&4.1 & 2.5&3.1 \\ 
0.1k& 5.7&7.6 & 2.2&2.9 & 1.5&1.9 \\ 
1k & 3.9&4.1 & 0.8&1.1 & 0.7&0.9 \\ 
2k & 2.4&3.1 & 0.6&0.8 & 0.6&0.7 \\ 
3k & 2.2&2.8 & 0.5&0.7 & 0.5&0.6 \\ 
4k & 1.9&2.4 & 0.5&0.6 & 0.4&0.6 \\ 
5k & 1.7&2.2 & 0.5&0.6 & 0.4&0.5 \\ 
\end{tabular}
\label{tab:Thermo}
\end{table} 

\subsection{Electron correlation}
To further assess the applicability of the $\Delta$-ML {\em Ansatz}, 
we modeled electron correlation energies,
essential for achieving chemical accuracy.~\cite{friesner2005ab}
Within post-HF theory, electron correlation energy can be defined as the difference 
between converged basis-set HF potential energy and its 
corresponding non-relativistic exact counterpart.~\cite{SzaboOstlund}
Evaluating the many-electron correlation energy at the post-HF level of theory 
requires substantial computational effort.
The computational complexity of the simplest post-HF method, 
second order perturbation theory (MP2), scales as $N_e^5$, where $N_e$ is 
the number of electrons.
The ``gold standard'' of quantum chemistry, CCSD(T), even scales as $N_e^7$. 
Revisiting the 6k constitutional isomers of 
C$_7$H$_{10}$O$_2$ (\ref{fig:density}), 
we have calculated the difference in the correlation energy part of 
molecular atomization energies for various correlated methods. 

\begin{figure}[hpt] 
\centering                                                                                
\includegraphics[width=8.8cm, angle=0.0]{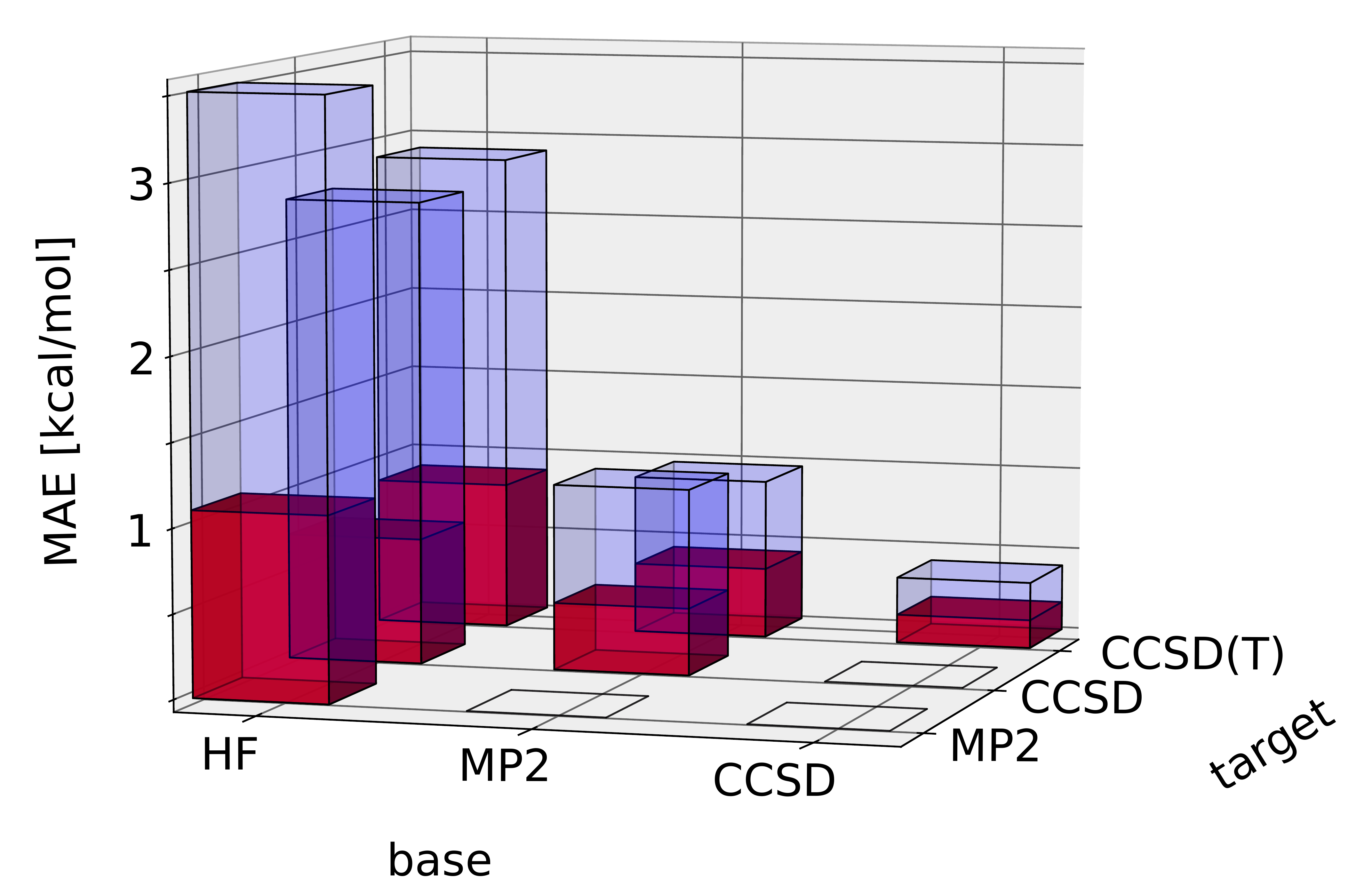}                                       
\caption{
Mean absolute error (MAE) [kcal/mol] of $E_b$ (blue bars) and 1k $\Delta_b^{t}$-ML 
model predictions (red bars) of atomization energies
for various combinations of increasingly correlated post-Hartree-Fock 
methods as target and baseline methods.
These values are obtained after subtracting the systematic average shift 
that exists between methods for the 6k isomers. 
See Supplementary Information for details.
}
\label{fig:electrons}                                                                          
\end{figure} 

For the 6k isomers MAEs, after accounting for  systematic shifts, 
are shown in \ref{fig:electrons} for various combinations of HF, MP2, CCSD, and CCSD(T),
along with the reduction of error due to 1k-$\Delta_{b}^{t}$ models.
Particularly noteworthy is the result for the 
$\Delta_{\rm HF}^{\rm CCSD(T)}$ model giving 
the total correlation energy (defined as difference
between HF and the exact result as approximated by CCSD(T))
between the least and most expensive method:
The MAE is reduced from $\sim$2.9 to  less than 1 kcal/mol.
Note that the 1k-$\Delta_{\rm HF}^{\rm MP2}$ 
model has a  larger MAE from MP2
than the MAE of the 1k-$\Delta_{\rm HF}^{\rm CCSD}$ from CCSD,
even though MP2 is more approximate in nature than CCSD. 
As such, the MP2 target, albeit less accurate and more approximate than
CCSD, appears to be a more complex function in chemical space. 

Remarkably, when using the 
1k-$\Delta_{\rm MP2}^{\rm CCSD}$
or
1k-$\Delta_{\rm MP2}^{\rm CCSD(T)}$
models the MAE  amounts in {\em both} cases to less than 0.5 kcal/mol. 
This suggests the possibility of the proverbial free lunch
modeling the more accurate and computationally more expensive
CCSD(T) level of theory with the same training set size and precision
as the less accurate and less expensive CCSD method.
The least approximate baseline, encoded by the 
1k-$\Delta_{\rm CCSD}^{\rm CCSD(T)}$ model, 
yields a chemically nearly negligible MAE from CCSD(T), $\sim$0.1 kcal/mol.
For larger training set sizes we have observed correlation energy errors to decay similarly 
to the error of DFT baseline models for thermodynamic properties.

\subsection{Applicability: Diastereomers of C$_7$H$_{10}$O$_2$}
We have tested the applicability of the
1k-$\Delta$-ML model, trained on 1k out of the 6k C$_7$H$_{10}$O$_2$
isomers in the GDB database, for the identification of the 
most stable diastereomers that can be generated from the parent isomers.
Such screening applications are highly relevant for spectroscopic 
or computational experiments aimed at the discovery and
characterization of competing reaction pathways, 
recently discussed for an ``{\em ab initio} nanoreactor''~\cite{Nanoreactor2014}.
More specifically, we applied the  
1k $\Delta_{\rm B3LYP}^{\rm G4MP2}$ model of atomization enthalpy 
at 298.15 K (\ref{tab:Thermo}), 
to screen all the 9868 unique and stable diastereomers 
resulting from inversion of atomic stereocenters in the 
original GDB set of 6095 constitutional isomers of C$_7$H$_{10}$O$_2$
(see Methods section).
For validation, we have randomly drawn 3k diastereomers and calculated
their computationally demanding G4MP2 enthalpies of atomization. 
The 1k-$\Delta_{\rm B3LYP}^{\rm G4MP2}$ model yields a MAE of
0.8 kcal/mol for these 3k diastereomers. 
We have chosen the DFT baseline for this exercise because cheaper baseline models, 
such as 5k $\Delta_{\rm PM7}^{\rm G4MP2}$ and $\Delta_{\rm PM7}^{\rm G4MP2^*}$ (\ref{fig:curve}),
exhibit less transferability when validated on the G4MP2 results for
the 3k diastereomers, namely MAEs of 3.5 and 2.8 kcal/mol, respectively.

\begin{figure*}[hpb] 
\centering                                                                                
\includegraphics[width=17cm, angle=0.0]{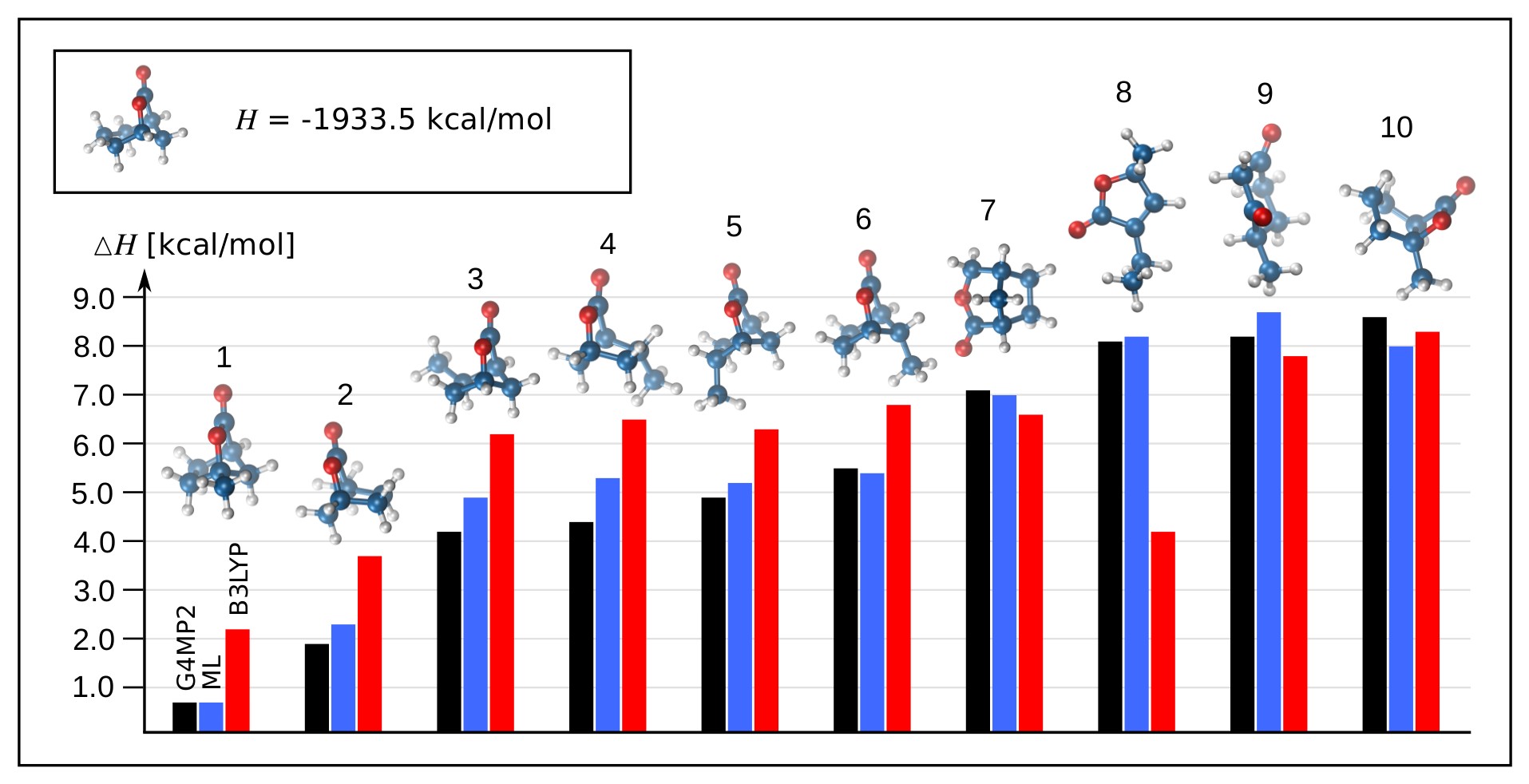}
\caption{         
Calculated reaction enthalpies at 298.15 K between the most stable molecule with 
C$_7$H$_{10}$O$_2$ stoichiometry (6-oxabicyclooctan-7-one, in inset, 
with atomization enthalpy -1933 kcal/mol), 
and its ten energetically closest isomers in increasing order, according to targetline
method G4MP2 calculated {\em a posteriori} (black).
1k $\Delta_{\rm B3LYP}^{\rm G4MP2}$ ML model predictions are given in blue. 
Baseline method B3LYP predictions are shown for comparison (red).
}                                                                                         
\label{fig:QMRC2D}                                                                          
\end{figure*}

Out of all the 10k diastereomers, the 1k-$\Delta_{\rm B3LYP}^{\rm G4MP2}$ model predicts
6-oxabicyclooctan-7-one, which is caprolactone with a methyl bridge between positions 1, and 5.
with an estimated atomization enthalpy $H$ of -1933.5 kcal/mol, to be the most stable isomer at ambient conditions. 
A validating G4MP2 calculation yielded the same number.
\ref{fig:QMRC2D} shows this molecule
along with its ten enthalpically closest isomers. 
These span a narrow energetic window of 9 kcal/mol, which
is sparse in comparison to the aforementioned 
100 molecules/kcal/mol energy density.
The six isomers for which $\Delta H<6$ kcal/mol
correspond to diastereomers of oxabicyclo[2.2.1]heptan-3-one, 
methylated at 1,4,5,5,6,7 positions, respectively. 
Isomers 3 ($\Delta H=4.3$ kcal/mol) and 4 ($\Delta H=4.5$ kcal/mol)
differ only by the chirality of the carbon atom at position 1.  
The next four high-lying isomers, although populating only a narrow $\Delta H$ range  
of 7--9 kcal/mol, exhibit very diverse chemical structures:
They include a cyclopentane fused with $\gamma$-valerolactone, a methyl, ethyl-substituted furanone, 
a methylated cyclo hexanedione, and
a cyclopentane fused with $\beta-$ propiolactone and methylated bridge atom framework. 
After identification of these isomers, we  calculated validating G4MP2 enthalpies
(\ref{fig:QMRC2D}).
The 1k-$\Delta_{\rm B3LYP}^{\rm G4MP2}$ ML model estimates the isomerization enthalpy
of these products with a maximal error of 0.6 kcal/mol for product 10.
The ML-model predictions agree with G4MP2 results calculated
{\em a posteriori} , and never exceed the
threshold of chemical accuracy (1 kcal/mol).

For comparison, \ref{fig:QMRC2D} also features estimates obtained 
from the DFT baseline method B3LYP, which is  popular among many computational 
as well as experimental chemists. 
While B3LYP would have predicted the same global minimum,
its reaction enthalpies can deviate substantially, and, 
sometimes  fail spectacularly (isomer 8).
It is interesting to note that the ML-model is apparently capable to 
reduce or increase the estimate depending on its baseline overshooting (isomers 1-6)
or underestimating (isomers 7-9). 
Only in the case of isomer 10, use of the ML-model would deteriorate the baseline's 
prediction error, albeit only from 0.3 to 0.5 kcal/mol.
We believe that such overall agreement of predicted reaction enthalpies 
with G4MP2 results obtained {\em a posteriori} strongly indicates
that the $\Delta$-ML {\em Ansatz} is capable to account for subtle errors made in 
the prediction of competitive chemical bonding---
at the baseline's computational cost (in this case DFT).

\subsection{Interpretation of the $\Delta$-Model:}

One can understand the trained corrections as follows:
The $\Delta_{\rm HF}^{\rm CCSD(T)}$ ML model of atomization energies
can be viewed as a ML model of  the correlation energy of atomization.
Likewise, when using atomization energies as baseline properties for free energies, 
and enthalpies, the difference in the resulting ML models cancels the baseline energy and
corresponds, after division by $T$, to the entropy of atomization
\begin{equation}
\left( \Delta^H_E  - \Delta^G_E \right)/T = S.
\label{eq:whatisdelta}
\end{equation}
Using a random  1k subset of the 6k C$_7$H$_{10}$O$_2$ isomers, we
have trained two ML models, one on $S$ of atomization at G4MP2 level of theory, taken as  
$(H-G)/T$ from Ref.~\onlinecite{DATAPAPER}, 
the other on $E_c$ at CCSD(T) level of theory, also from Ref.~\onlinecite{DATAPAPER}.
Computationally efficient PM7 
equilibrium geometries have been used for training, testing, and predictions. 

\begin{figure}[hpt]
\centering
\includegraphics[width=8.8cm, angle=0.0]{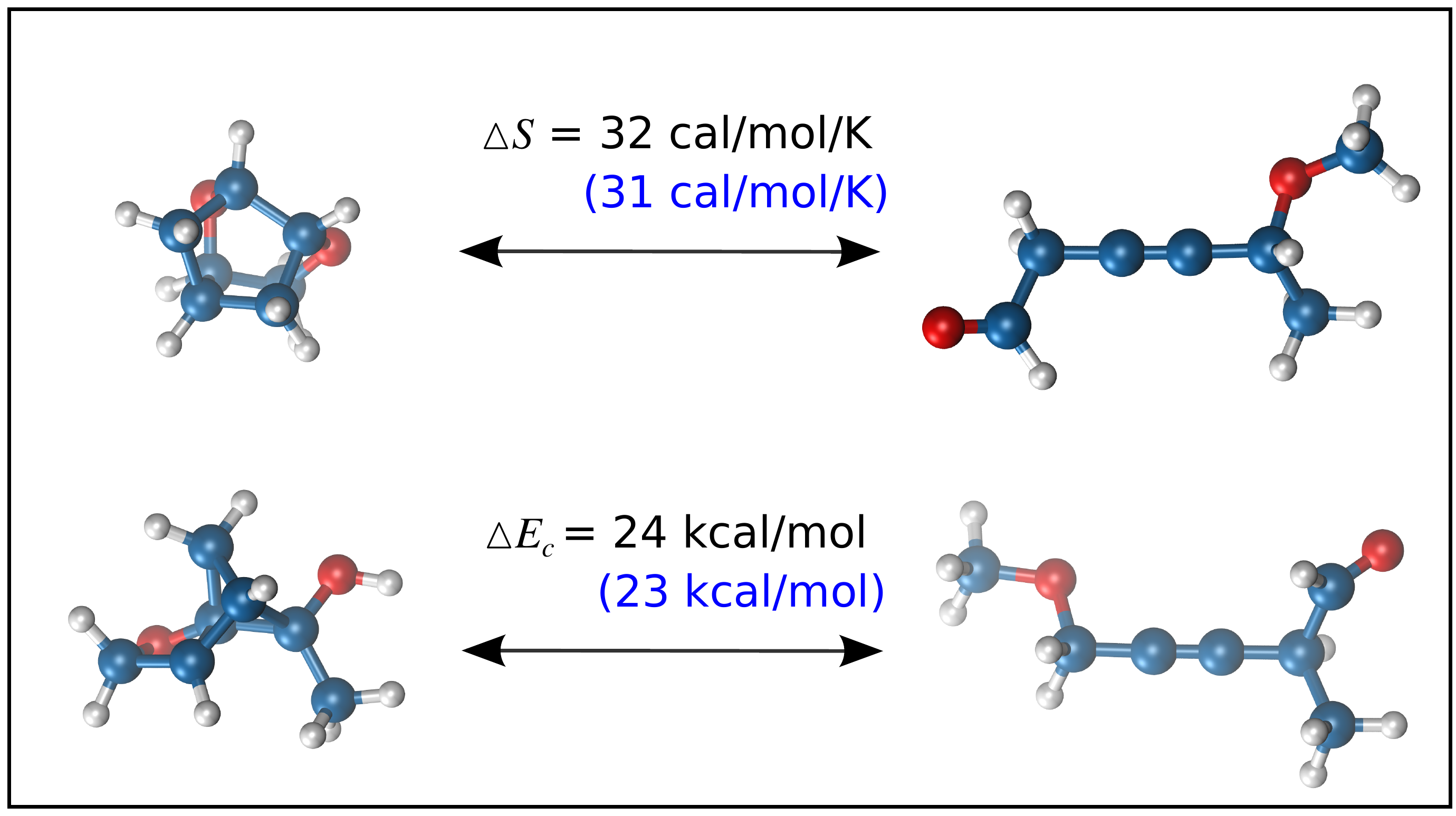}
\caption{
Molecular pairs with maximal reaction entropy (top) and electron
correlation energy (bottom) out of the 10k stable diastereomers of C$_7$H$_{10}$O$_2$.
Reaction properties have been estimated using a
1k-$\Delta_{\rm B3LYP}^{\rm G4MP2}$-ML model for the entropy and a
1k-$\Delta_{\rm HF}^{\rm CCSD(T)}$-ML model  for the correlation energy (black).
Corresponding reference values calculated for validation are given in parentheses (blue).
}
\label{fig:QMRCD}
\end{figure}

We have reapplied the resulting 1k models to screen 
the aforementioned 10k diastereomers for those molecular pairs
which exhibit maximal isomerization entropies $\Delta S$ and correlation energies $\Delta E_c$.
Structures of the molecules with extreme $S$ and $E_c$ are shown in \ref{fig:QMRCD}.
The molecular pair with maximal $\Delta S$ is consistent with chemical intuition:
The lowest entropy isomer, 2,5-dioxatricyclononane, has a cage-like structure and is 
very compact, bearing some resemblance to adamantane, and void of any conformational degree of freedom.
By contrast, the molecule with largest entropy, 5-methoxyhex-3-ynal, possesses multiple conformational
degrees of freedom, made possible through the occurrence of a double and a triple bond
that consume the valences otherwise accessible for ring or cage formation. 
The resulting $\Delta S$ estimated by the ML model deviates from the reference G4MP2 value by 
only 1 cal/mol/K.
Quantitative rationalization of the trend in $E_c$ is less obvious, 
owing to its origin in electronic many-body effects.
However, $E_c$ can be intuitively related to the number of interacting electron pairs.
This number is small when the molecule is long since
according to the nearsightedness principle~\cite{Nearsightedness}, electrons localized
on one end of the molecule interact less with those from the other end.
In compact molecules by contrast, more electrons can interact,  hence the number of
interacting electron pairs, and consequently the magnitude of $E_c$, are large.
After screening of diastereomers using our $E_c$ model
we found among 10k diastereomers  6-methyl-2-oxatricycloheptan-6-ol
and 5-methoxy-2-methylpent-3-ynal to have the maximal reaction electron correlation energy, see \ref{fig:QMRCD}.
The maximal electron correlation energy difference, 24 kcal/mol, deviates from validating
CCSD(T) reference results by 1 kcal/mol. 
Both molecules confirm  intuition: The most compact molecule with few  
degrees of freedom exhibits maximal correlation while the most elongated 
corresponds to the least amount of correlation energy.

\begin{figure}[hpt]
\centering
\includegraphics[width=8.8cm, angle=0.0, scale=1]{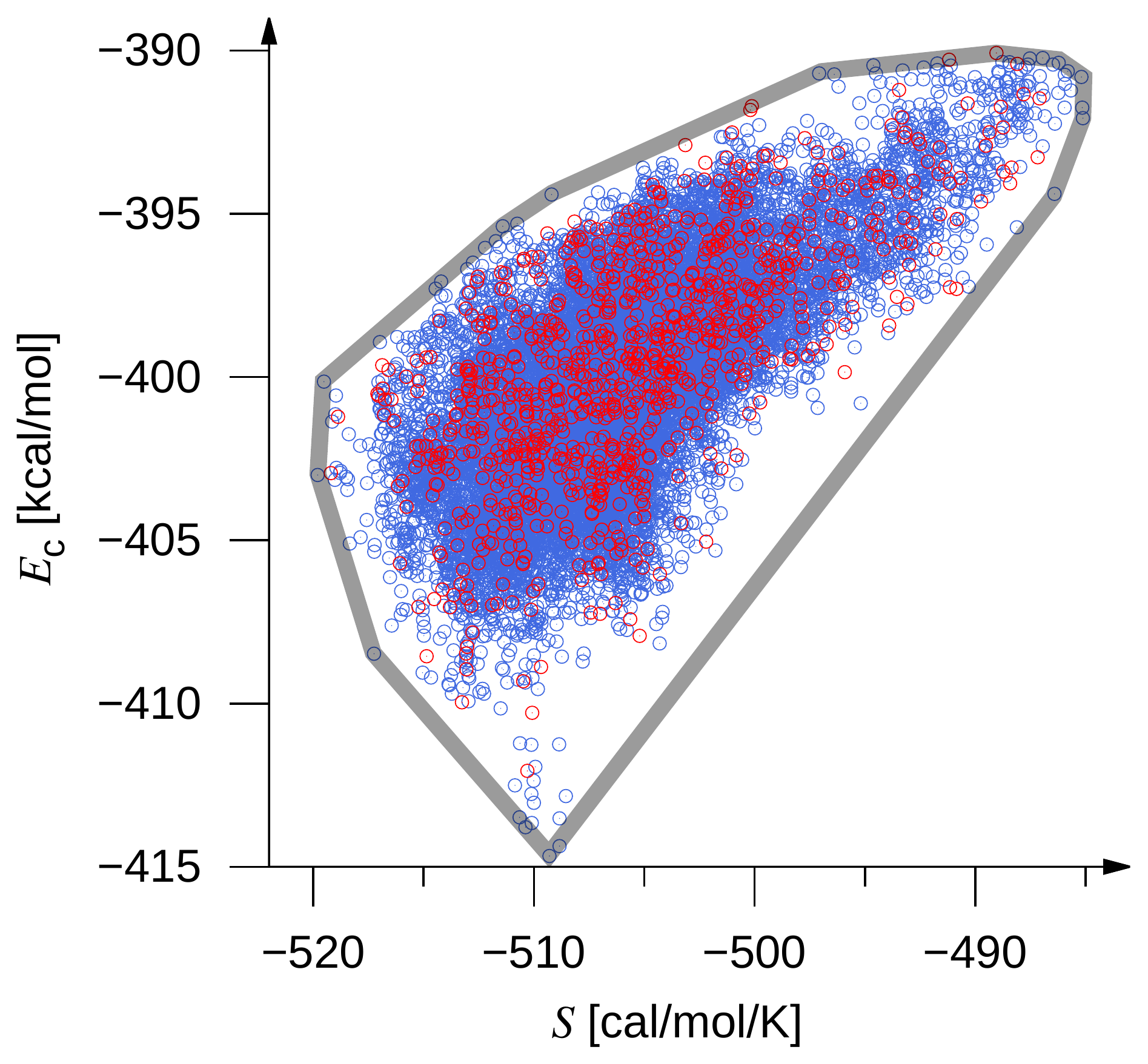}
\caption{
Scatter plot of ML-model predicted entropy of atomization at $T$ = 298.15 K versus 
ML-model predicted electron correlation contribution to atomization energy
for the 16k stable out-of-sample diastereomers of C$_7$H$_{10}$O$_2$.
All predictions are made using ML models trained on 1k molecules randomly drawn
from set of 6k parent isomers. 
Training data is shown in red.
Linear regression yields $E_c \approx~0.445 \times TS~-175.211$ [kcal/mol].
Pareto fronts are indicated by convex hull shown in gray. 
}
\label{fig:pareto}
\end{figure}

In above discussion we notice that for both pairs with maximal difference in electron correlation
and in entropy, respectively, similar observations hold: Compactness/extension appears to maximize
the difference in both cases.
This observation raises the question whether $S$, which arises from the vibrational partition function,
and $E_c$, due to electronic many-body effects, are interdependent.
An underlying relationship could aid not only in pin-pointing molecules
that pose interesting challenges for benchmarking approximate electronic theories
but might even lead to semi-quantitative estimations of $E_c$ via $S$.
Thermal molecular properties, such as heat-capacities, 
could be linked directly to their electronic structure.
To  elucidate their potential relationship,  
\ref{fig:pareto} shows
a scatter plot of the model-predicted atomization entropy and correlation energies of
the 10k diastereomers, as well as the 6k parent isomers of C$_7$H$_{10}$O$_2$.
Albeit hardly quantitative, a qualitative interdependence is revealed, 
suggesting a molecular analogue to phonon-electron coupling phenomena in solids. 

Such a dependency might serve the construction of rough structure property relationships for the 
filtering of compounds using one property as a mutual descriptor for the other.
Furthermore, this relationship could possibly impose severe 
constraints on how freely $S$ and $E_c$ can be varied independently 
within multi-objective property optimization procedures in chemical compound space.
To further illustrate this point, \ref{fig:pareto} also highlights  corresponding Pareto fronts. 
Note, for example, that while \ref{fig:QMRCD} displays the pairs that maximize the
vertical ($E_c$) or horizontal ($S$) axis in  \ref{fig:pareto}, the molecular pair that
simultaneously maximizes both differs.
Other molecules, such as bullvalene, also happen to fall onto the same  linear relationship. 
However, for organic molecules with very different sizes, taken from the 134k GDB-9 dataset, 
this linear trend breaks down.
As such, it might still require normalization by number of atoms or electrons to hold in general.

We finally note that arriving at these observations exclusively via high-throughput \emph{ab initio}
computations would have required $N_e^7$-scaling G4MP2 calculations for all the 10k diastereomers
with an estimated need for compute time of $\sim$20 CPU years.
The PM7 baseline predictions, by contrast, required only $\sim$1 CPU day for all geometry relaxations, 
and the remaining deviation from target properties G4MP2-$S$ and CCSD(T)-$E_c$ is given
instantaneously by the ML correction.

\subsection{Thermochemistry for 134 kilo organic molecules}
When dealing with hundreds of thousands of molecules one typically assumes that it 
is not necessary to achieve chemical accuracy for all of them. 
Instead, hierarchical procedures where less accurate but computationally
more efficient methods, such as DFT, filter out the most relevant compounds which subsequently 
can be studied using more accurate and computationally more demanding methods, such as G4MP2.
DFT calculations, however, are ordinarily too expensive to be used for filtering hundreds 
of thousands of molecules. 
We have investigated whether the $\Delta$-approach can be used for DFT-quality filtering 
at the computational cost of a semi-empirical quantum chemistry calculation.
Specifically, based on PM7 baselines we have predicted DFT targetline 
enthalpies and entropies of atomization.
To more systematically assess transferability, we have trained a 1k and 10k 
training set drawn at random from the 
nearly 134k organic molecules containing up to nine C, N, O, or F atoms
(published as GDB-9 in Ref.~\onlinecite{DATAPAPER}).
For subsequent validation, we have used the remaining 133k and 124k molecules, respectively. 

\begin{figure}[hpb]
\centering                                                  
\includegraphics[width=8.8cm, angle=0.0, scale=1]{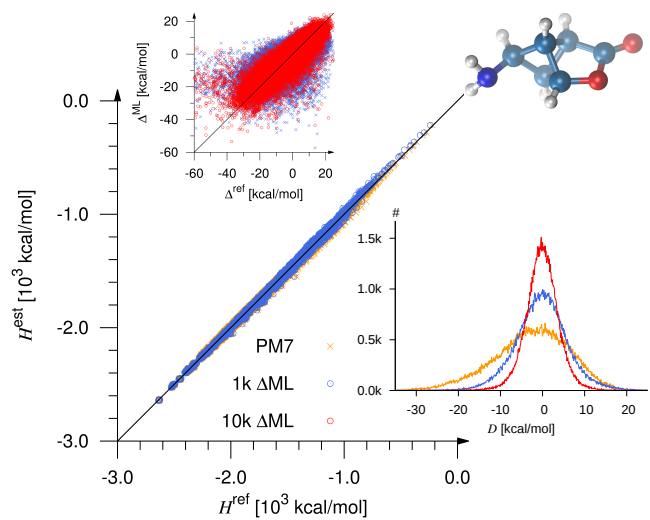}
\caption{
Scatter plot of predicted atomization enthalpies $H$ at 298.15 K for 124k out-of-sample 
GDB-9 molecules \cite{GDB17}.
Estimated $H$ is plotted for PM7 (yellow, MAE = 7.2~kcal/mol) and 1k (blue) and 10k (red)
$\Delta_{\rm PM7}^{\rm B3LYP}$-ML models
(MAE=4.8 and 3.0~kcal/mol, respectively) versus reference B3LYP values. 
The left side inset shows the ML contribution to the estimated energy differences between 
PM7 and reference B3LYP enthalpies, $\Delta^{\rm est}$,
for 1k (blue) and 10k (red) models versus reference difference, $\Delta^{\rm ref}$.
The right side inset shows the error distribution around B3LYP enthalpies 
for PM7, 1k, and 10k models, respectively. The most extreme outlier (top, right),
7-amino-3-oxatricycloheptan-4-one, has error
86, 74, and 58 kcal/mol for PM7, 1k, and 10k models, respectively.
}
\label{fig:scatter}                                                                          
\end{figure}

On average, PM7 enthalpies of atomization deviate from B3LYP by 7.2~kcal/mol. 
For a randomly drawn training set of 1k molecules, 1k-$\Delta_{\rm PM7}^{\rm B3LYP}$-ML
predicts B3LYP enthalpies of the 133k additional (out-of-sample) molecules with an MAE of 4.8~kcal/mol. 
Increasing the number of training molecules to 10k leads to an improved MAE of 3.0~kcal/mol,  
as measured for the remaining 124k out-of-sample molecules.
We note that such a predictive accuracy places the 
10k-$\Delta_{\rm PM7}^{\rm B3LYP}$-ML model on par with generalized 
gradient approximated (GGA) or even hybrid 
DFT~\cite{ChemistsGuidetoDFT, DFTreview-CohenYang2012cr}---at the computational cost of PM7.
\ref{fig:scatter} features the corresponding scatter plot of actual versus 
predicted B3LYP enthalpies of atomization. 
The lower right inset shows that the baseline's systematic underestimation, 
as well as its skew, has been removed already by the 1k-$\Delta_{\rm PM7}^{\rm B3LYP}$-ML model.
The error distribution contracts further as the training set size is increased to 10k.
The upper left inset scatter plot illustrates the importance of having a baseline, 
the pure ML contribution being far from perfect correlation. 
The molecular structure on is the most extreme outlier, PM7 
underestimates its stability by 86.0~kcal/mol. 
Encouragingly, 1k and 10k-$\Delta_{\rm PM7}^{\rm B3LYP}$-ML models reduce the error for this 
outlier to 73.9 and 58.0~kcal/mol, respectively.

\begin{figure*}[hpt]                                                                       
\centering                                                  
\includegraphics[width=17cm, angle=0.0, scale=1.0]{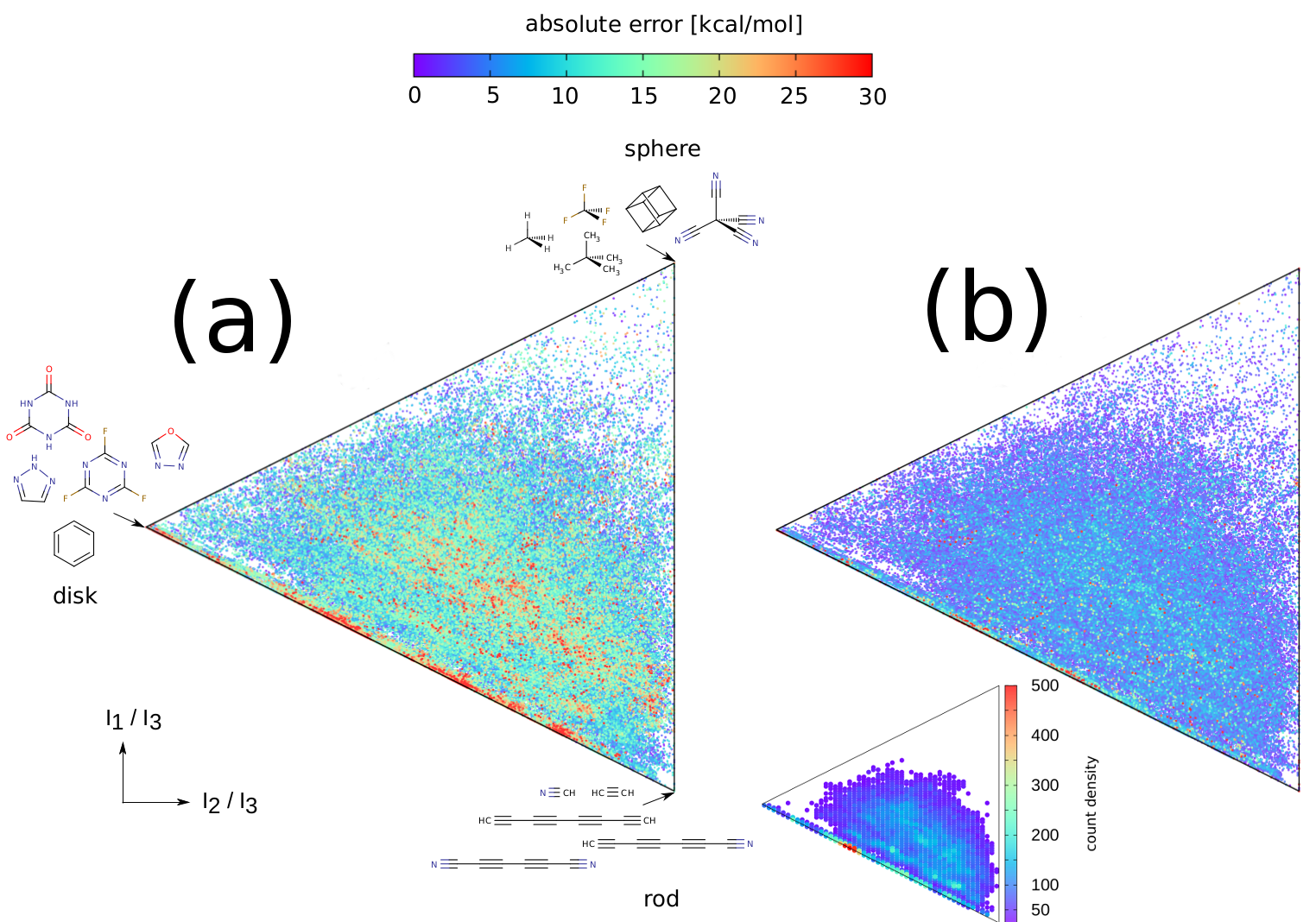}
\caption{
Shape distribution of color-coded absolute deviations between predicted and B3LYP reference 
atomization enthalpies of 
the 134k organic molecules with up to nine atoms (not counting hydrogens)
in the GDB-17 data set~\cite{GDB17}.
Vertical and horizontal axes correspond to 
normalized principal moments of inertia
$I_1/I_3$ and $I_2/I_3$, respectively, 
with $I_1 \le I_2 \le I_3$. 
(a) Large PM7 errors ($>$ 25 kcal/mol) are predominantly present for 
geometries on the rod-disk edge, and in the center of the triangle.
Corners indicate the geometrical shape of molecules with 
molecular drawings corresponding to examples of linear ($I_1=0$, $I_2=I_3$), 
planar ($2I_1=2I_2=I_3$), and spherical ($I_1=I_2=I_3$) cases.
(b) 10k-$\Delta_{\rm PM7}^{\rm B3LYP}$-ML errors are systematically
smaller, some outliers at the rod-disk edge persist.
A heatmap of the molecular data density is shown for the same coordinate system in the inset below (b).
}
\label{fig:134k}                                                                          
\end{figure*} 

We have also analyzed the effect of molecular geometry.
It is well known that the faithfulness of common quantum chemical methods
can alter drastically when changing the geometry of the molecule.
Stretching chemical bonds, for instance, can lead to severe
errors, even for methods that predict the energy minimum perfectly well.
To systematically assess the effect of geometry, 
we compare the predicted B3LYP atomization enthalpies for PM7 
and 10k-$\Delta_{\rm PM7}^{\rm B3LYP}$ for all 134k molecules,
as a function of their normalized principal moments of inertia.
\ref{fig:134k} displays the resulting deviation from B3LYP, 
spanned by molecular geometry (rod, disk, or sphere-like).
While PM7 has particularly strong deviations ($\sim$ 20 kcal/mol) 
on the linear to planar edge, 
as well as close to the lower part of the linear to spherical edge, 
use of the ML correction homogeneously quenches the error throughout
the triangle into the 5 kcal/mol error window, with very few 20 kcal/mol
outliers persisting on the rod-disk edge. Note that due to the non-uniqueness of the
moments of inertia, error heatmaps in \ref{fig:134k} of many molecules superimpose each other
in increasing order.  To avoid possible mis-interpretations,
the inset with a heat-map of data density provides a means to visually
normalize the error heatmaps.

Regarding the computational speed-up, we note that on a typical CPU, 
a single $\Delta_{\rm PM7}^{\rm B3LYP}$-ML evaluation                     
requires no more than 10 seconds for the largest molecule in GDB-9.
Thus, screening of the entire set of 134k molecules has consumed less than 2 CPU weeks.
By contrast, the average computational cost for obtaining a B3LYP atomization enthalpy
amounts to roughly 1 CPU hour per molecule, implying 15 CPU years for DFT 
based screening of the 134k molecules.

\subsection{Conclusions}

We have introduced a composite quantum chemistry/machine learning approach.
It combines approximate but fast legacy quantum chemical approximations
with modern big data-based machine learning estimates trained on expensive and accurate
reference results throughout chemical space.
We have shown that the $\Delta$-ML model can be used to study other, out-of-sample molecules, not part of training.
Effectively one can reach the accuracy of high-level quantum chemistry at
a dramatically lower computational burden which
is dominated by the employed baseline method, such as semi-empirical quantum-chemistry (PM7), HF, or DFT.
Mere reparameterization of the baseline method's global parameters for a given training
set does not suffice, yielding measurable advantage only for very small and selected training and test sets.
Alternative molecular representations, however, could still lead to faster convergence to chemical accuracy. 
Similar learning rates with respect to training set size among all model-combinations, merely differing by off-set, 
suggest that even very approximate and computationally inexpensive baseline models can be used,
provided access to sufficiently large training sets. 
For chemically diverse sets of organic molecules we have presented numerical evidence that
chemically accurate molecular thermochemistry predictions can be made at a computational cost reduced by several orders 
of magnitude when compared to the reference method for new out-of-sample molecules. 

For the most stable isomer in the set of 10k diastereomers generated from all
6k molecules with C$_7$H$_{10}$O$_2$ stoichiometry in GDB-17~\cite{GDB17},
we have demonstrated how to identify the ten most competitive reaction isomers.
For the same diastereomers we also identified a qualitative dependency between
entropy and correlation energy of atomization, suggesting a molecular equivalent
of electron-phonon coupling.
Finally, we have presented evidence for the transferability of the $\Delta$-ML model by
reducing the error of semi-empirical quantum chemistry method from 7.2 kcal/mol
to the error of generalized gradient approximated ($\sim$ 5 kcal/mol) 
or hybrid density functional theory ($\sim$ 3 kcal/mol)
for over hundred thousand organic molecules using less
than 1 and 10\% of them for training, respectively. 

We believe the high predictive accuracy to be due to the
fact that approximate theories already capture the most important contributions to chemical energetics. 
The remaining deviations from the reference results are typically smaller, possibly also smoother, 
and prove to be more amenable to statistically trained ML models. 
Overall, our results suggest that the $\Delta$-ML-model represents an attractive strategy for 
augmenting legacy quantum chemistry with modern big data driven ML models. 
For future studies, this strategy might also offer substantial improvements to predictive
accuracy of  other properties such as heat capacities, non-adiabatic energy corrections, 
barriers of elementary reaction steps, optical properties, atomic forces for molecular dynamics calculations, 
molecule specific parameters  for semi-empirical theories, or electronic excitations.

\section{Methods}
\subsection{Molecular datasets}
We have considered four sets of organic molecules. 
The first set has been used for preliminary testing of the {\em Ansatz}, 
and consists of the 7211 (7k) organic molecules and HOMO/LUMO eigenvalues and 
molecular polarizabilities at different levels of theory as
published in Ref.~\onlinecite{Montavon2013}.
The second set contains 133885 (134k) molecules with up to 
9 heavy atoms (C, O, N, F, not counting H) in the
universe of small organic molecules ``GDB''~\cite{GDB17} for which
we calculated and published semi-empirical (PM7) and density functional theory 
(B3LYP)-based thermochemical properties such
as enthalpies and entropies of atomization~\cite{DATAPAPER}.
The diversity of this set is shown in \ref{fig:134k}. 
We note at this point that in $\Delta$-ML models other baseline methods, 
such as extended H\"uckel, tight-binding DFT~\cite{ElstnerDFTTB}, 
OM2~\cite{weber2000orthogonalization}, or AM05~\cite{AnnsSurfacefctlPRB2005} 
could have been used just as well.
The third set corresponds to a subset of the second set: 
For 6095 (6k) constitutional isomers of C$_7$H$_{10}$O$_2$
we calculated the same thermochemical properties at 
significantly more sophisticated and computationally demanding level of theory,
widely considered to be of ``chemical accuracy'' ($\sim$1~kcal/mol).
Also this set has been published in Ref.~\onlinecite{DATAPAPER}.
Finally, the versatility of this method is assessed for 
a fourth set of molecules, consisting of 9868 (10k) stable diastereomers that are not part of the GDB
universe, and have been obtained by inverting all atomic stereocenters 
in the aforementioned third set of 6k C$_7$H$_{10}$O$_2$-isomers. 
This dataset is a part of this publication, and is available on the authors' homepage. 

\subsection{Computational details}~
From Ref.~\onlinecite{GDB17}, we obtained all SMILES~\cite{w1988} strings for molecules with up to nine heavy atoms. 
We then excluded cations, anions, and molecules containing S, Br, Cl, or I, arriving at 133885 molecules. 
This data is presented and analyzed in more depth in Ref.~\onlinecite{DATAPAPER}. 
Cartesian coordinates for the subset of 6095 isomers of C$_7$H$_{10}$O$_2$ were 
determined by parsing the corresponding SMILES strings using Openbabel software~\cite{Openbabel1}, 
followed by a consistency check using the CORINA code~\cite{CORINA}. 
Structures of 9868 non-enantiomeric stable diastereomers were obtained 
through inversions of chiral C atoms in the SMILES strings followed
by conversion to Cartesian coordinates using CORINA.
To verify that all theoretical methods preserved topology and chirality, we transformed the Cartesian coordinates back to 
SMILES, and
InChI strings using Openbabel. 
Using these initial structures, we carried out geometry relaxations at the PM7~\cite{PM7}
semi-empirical level of theory using MOPAC~\cite{MOPAC2012}. 
We used the PM7 equilibrium coordinates as initial geometries and performed DFT and G4MP2 geometry 
calculations using Gaussian09~\cite{Gaussian09D01}. 
For DFT calculations, we employed the Gaussian basis set 6-31G(2df,p) 
which is also used in the G4MP2 calculations in combination with the DFT 
method B3LYP \cite{B3LYP}, for geometry relaxation and frequency computations. We
used the same basis set also in the GGA-PBE~\cite{PBE} calculations.
G4MP2 employs harmonic oscillator and rigid rotor approximations
to estimate the entropy of nuclear degrees of freedom \cite{G4MP2}. 
At all levels of theory, we performed harmonic vibrational analysis for all molecules to confirm 
that the predicted equilibrium structures were local minima on the potential energy 
surface. HF, MP2, CCSD, CCSD(T) energies  have been computed with the basis set 6-31G(d) as a part of G4MP2. 
Further technical details regarding all quantum chemistry data, 
including convergence thresholds employed, are given in Ref.~\onlinecite{DATAPAPER}.

\section{Acknowledgement}
The authors thank J.-L.~Reymond for discussions and GDB SMILES, 
J.~Stewart and C.~H.~Schwab for providing trial licenses for the packages 
MOPAC and Corina, respectively. 
A.~De Vita, K-R.~M{\"u}ller, A.~Tkatchenko, and P.~W.~Ayers are acknowledged for discussions. 
Most calculations were performed at sciCORE (http://scicore.unibas.ch/) 
scientific computing core facility at University of Basel.
This research also used resources of the Argonne Leadership Computing Facility at Argonne National 
Laboratory, which is supported by the Office of Science of the U.S. DOE under 
contract DE-AC02-06CH11357.  
This work was funded by the Swiss National Science foundation (No.~PP00P2\_138932).  



\end{document}